\documentclass{JHEP3} % 10pt is ignored!
\usepackage{amsmath,amssymb}
\usepackage{type1cm}
\usepackage{epsfig}
\usepackage{bbm}
\usepackage{ifthen}

\DeclareMathOperator{\tr}{tr}

\DeclareMathOperator{\Pf}{Pf}
\DeclareMathOperator{\SU}{SU}

\DeclareMathOperator{\U}{U}

\DeclareMathOperator{\Arg}{Arg}
\newcommand{\rmd}{{\rm d}}

\newcommand\fverb{\setbox\pippobox=\hbox\bgroup\verb}
\newcommand\fverbdo{\egroup\medskip\noindent%
                        \fbox{\unhbox\pippobox}\ }
\newcommand\fverbit{\egroup\item[\fbox{\unhbox\pippobox}]}
\newbox\pippobox
\newcommand{\llangle}{\left\langle\!\left\langle}
\newcommand{\rrangle}{\right\rangle\!\right\rangle}
\newcounter{private}
\title{Two-dimensional $\mathcal{N}=(2,2)$ super Yang-Mills theory on
computer}

\author{Hiroshi Suzuki\\
Theoretical Physics Laboratory, RIKEN, Wako 2-1, Saitama 351-0198, Japan\\
E-mail: \email{hsuzuki@riken.jp}}

\received{\today}               %%
\accepted{\today}               %% These are for published papers.
\preprint{\today}
\preprint{RIKEN-TH-103}

\abstract{We carry out preliminary numerical study of Sugino's lattice
formulation~\cite{Sugino:2004qd,Sugino:2004qdf} of the two-dimensional
$\mathcal{N}=(2,2)$ super Yang-Mills theory (2d $\mathcal{N}=(2,2)$ SYM) with
the gauge group~$\SU(2)$. The effect of dynamical fermions is included by
re-weighting a quenched ensemble by the pfaffian factor. It appears that the
complex phase of the pfaffian due to lattice artifacts and flat directions of
the classical potential are not problematic in Monte Carlo simulation. Various
one-point supersymmetric Ward-Takahashi (WT) identities are examined for
lattice spacings up to~$a=0.5/g$ with the fixed physical lattice
size~$L=4.0/g$, where $g$~denotes the gauge coupling constant in two
dimensions. WT identities implied by an exact fermionic symmetry of the
formulation are confirmed in fair accuracy and, for most of these identities,
the quantum effect of dynamical fermions is clearly observed. For WT identities
expected only in the continuum limit, the results seem to be consistent with
the behavior expected from supersymmetry, although we do not see clear
distintion from the quenched simulation. We measure also the expectation values
of renormalized gauge-invariant bi-linear operators of scalar fields.
}


\begin{document} 

\maketitle %%%%%%%%%% THIS IS IGNORED %%%%%%%%%%%

\section{Introduction}
It will be very exciting if non-perturbative question in supersymmetric gauge
theories (such as possibility of spontaneous breaking of supersymmetry) can be
studied numerically at one's will. Despite the great efforts being made towards
numerical study of the four-dimensional $\mathcal{N}=1$ super Yang-Mills theory
(4d $\mathcal{N}=1$ SYM)~\cite{Curci:1986sm}--\cite{Farchioni:2004fy}, so far
no conclusive evidence of a restoration of supersymmetry in the continuum limit
has been observed. For recent reviews on lattice formulation of supersymmetric
theories, see refs.~\cite{Kaplan:2003uh,Feo:2004kx,Giedt:2006pd}. Under this
situation, to test various ideas, it seems useful to examine lower dimensional
supersymmetric gauge theories in great detail, which have much simpler
ultraviolet (UV) structure and for which it is relatively easy to accumulate
high statistics in Monte Carlo simulation.

In this paper, we report the results of our small-scale Monte Carlo study of
lattice formulation of the 2d $\mathcal{N}=(2,2)$ SYM, proposed by
Sugino~\cite{Sugino:2004qd,Sugino:2004qdf}. For this and similar
lower-dimensional supersymmetric gauge theories, many other proposals and
studies on possible lattice formulation
exist~\cite{Kaplan:2002wv}--\cite{Takimi:2007nn}.
(See also ref.~\cite{Matsumura:1995kw}
for studies based on the supersymmetric discrete light-cone quantization.)
The advantage of the formulation of refs.~\cite{Sugino:2004qd,Sugino:2004qdf}
is that a fermionic symmetry, associated with one of four supercharges of the
target theory, is manifestly preserved even with finite lattice spacings and
finite volume. Full supersymmetry is expected to be restored in the continuum
limit. Possible disadvantage of the formulation, on the other hand, is that the
pfaffian resulting from the integration over fermionic fields is generally
complex,\footnote{To avoid this point is one of motivations of the proposal of
ref.~\cite{Suzuki:2005dx}.} although the complex phase is expected to be
irrelevant in the continuum limit, as the corresponding pfaffian in the target
theory is real and positive semi-definite.

In our simulation, we include the effect of dynamical fermions by re-weighting.
That is, in taking a statistical average, a quenched ensemble is re-weighted by
the factor of pfaffian. With parameters and statistics of our Monte Carlo
simulation, it appears that the complex phase of the pfaffian and flat
directions of the classical potential (which might imply subtlety in the
integration over scalar fields) are not problematic. The parameters of our
simulation correspond to lattice spacings up to~$a=0.5/g$ with the fixed
physical lattice size~$L=4.0/g$, where $g$ denotes the gauge coupling constant
in two dimensions.

In this paper, we mainly study one-point supersymmetric bare WT identities.
These are precisely WT identities numerically analysed by
Catterall~\cite{Catterall:2006jw} on the basis of his lattice formulation of
the 2d $\mathcal{N}=(2,2)$ SYM~\cite{Catterall:2004np}. In our numerical
simulation, WT identities implied by the exact fermionic symmetry of the
formulation are reproduced in fair accuracy and, for most of these identities,
we clearly observe the quantum effect of dynamical fermions. For WT identities
expected only in the continuum limit, the results seem to be consistent with the behavior expected by supersymmetry, although we
do not see clear distinction from the quenched (i.e., non supersymmetric)
simulation. We measure also the expectation values of renormalized
gauge-invariant bi-linear operators of scalar fields to illustrate how this
kind of numerical study would be useful.

In section~2, we briefly review Sugino's formulation of the 2d
$\mathcal{N}=(2,2)$ SYM mainly to fix our notation. Some remarks are made on
the continuum limit. In section~3, the results of our Monte Carlo simulation
are reported. In section~3.1, we explain our simulation algorithm and related
matters. In section~3.2, one-point WT identities are studied. In section~3.3,
expectation values of gauge-invariant bi-linear operators of scalar fields are
studied. Section~4 is devoted to conclusion. Throughout this paper, the gauge
group is assumed to be $\SU(N_c)$ and our simulation has been done only for
$\SU(2)$.

\section{Sugino's lattice formulation of the 2d $\mathcal{N}=(2,2)$ SYM}

\subsection{Topological field theoretical form of the continuum target theory}
This lattice formulation starts with the fact that the (euclidean) action
of the 2d $\mathcal{N}=(2,2)$ SYM can be written in the form of the topological
field theory~\cite{Witten:1988ze}\footnote{The conventional form of the action
of the 2d $\mathcal{N}=(2,2)$ SYM, for example, eq.~(2.7) of
ref.~\cite{Suzuki:2005dx}, is reproduced by the following substitution
\begin{align}
   &A_\mu\to -ig\sum_aA_\mu^aT^a,\qquad
   \phi\to-ig\sum_a(\varphi^a+i\phi^a)T^a,\qquad
   \overline\phi\to-ig\sum_a(\varphi^a-i\phi^a)T^a,
\nonumber\\
   &\psi_0\to-ig\sum_a(-i\psi_1^a+i\psi_2^a
   -\overline\psi_1^a+\overline\psi_2^a)T^a/2,\qquad
   \psi_1\to-ig\sum_a(\psi_1^a-\psi_2^a
   +i\overline\psi_1^a-i\overline\psi_2^a)T^a/2,
\nonumber\\
   &\chi\to-ig\sum_a(-\psi_1^a-\psi_2^a
   -i\overline\psi_1^a-i\overline\psi_2^a)T^a/2,\qquad
   \eta\to-ig\sum_a(i\psi_1^a+i\psi_2^a
   +\overline\psi_1^a+\overline\psi_2^a)T^a/2,
\label{twoxone}
\end{align}
where $T^a$ are anti-hermitian generators of $\SU(N_c)$ normalized as
$\tr\{T^aT^b\}=-(1/2)\delta_{ab}$ and the index~$a$ runs from 1 to $N_c^2-1$.}
\begin{align}
   S_{\text{continuum}}&=\frac{1}{g^2}
   \int\rmd^2x\,\tr\biggl\{
   \frac{1}{4}[\phi,\overline\phi]^2+H^2-iH\Phi
   +D_\mu\phi D_\mu\overline\phi
\nonumber\\
   &\qquad\qquad\qquad\qquad{}
   -\frac{1}{4}\eta[\phi,\eta]-\chi[\phi,\chi]
   +\psi_\mu[\overline\phi,\psi_\mu]
   +i\chi Q\Phi+i\psi_\mu D_\mu\eta
   \biggr\},
\label{twoxtwo}
\end{align}
where all fields are $\SU(N_c)$ Lie algebra valued and scalar fields $\phi$
and~$\overline\phi$ are combinations of two real scalar fields, $\phi=X_2+iX_3$
and $\overline\phi=X_2-iX_3$, respectively. $\Phi=2F_{01}$ is the field strength
in two dimensions $F_{01}=\partial_0A_1-\partial_1A_0+i[A_0,A_1]$. The covariant
derivatives $D_\mu$ are defined with respect to the adjoint representation
$D_\mu\varphi=\partial_\mu\varphi+i[A_\mu,\varphi]$ for any field~$\varphi$. The
index~$\mu$ runs over 0 and~1. Note that, in the above convention, the bosonic
fields $A_\mu$, $\phi$ and~$\overline\phi$ have the mass dimension~1 and the
fermionic fields $\psi_\mu$, $\chi$ and~$\eta$ have the mass dimension~$3/2$,
because the gauge coupling constant in two dimensions~$g$ has the mass
dimension~1.

In eq.~(\ref{twoxtwo}), $Q$ is a BRST-like transformation in the topological
field theory (that is a particular linear combination of super-transformations
in the original SYM theory) and is defined by
\begin{align}
   &QA_\mu=\psi_\mu,&&Q\psi_\mu=iD_\mu\phi,
\nonumber\\
   &Q\phi=0,&&
\nonumber\\
   &Q\chi=H,&&QH=[\phi,\chi],
\nonumber\\
   &Q\overline\phi=\eta,&&Q\eta=[\phi,\overline\phi].
\label{twoxthree}
\end{align}
The salient feature of this transformation is that its square~$Q^2$ is an
infinitesimal gauge transformation with the transformation parameter~$\phi$.
Therefore, $Q$ is nilpotent $Q^2=0$ when acting on gauge invariant quantities.
Moreover, the action can be expressed as a $Q$-exact form:
\begin{equation}
   S_{\text{continuum}}=Q\frac{1}{g^2}
   \int\rmd^2x\,\tr\left\{
   \frac{1}{4}\eta[\phi,\overline\phi]-i\chi\Phi+\chi H
   -i\psi_\mu D_\mu\overline\phi
   \right\}.
\label{twoxfour}
\end{equation}
In this form, the $Q$-invariance of the action is manifest. Then the
idea\footnote{See ref.~\cite{Catterall:2001fr}.} is to construct a lattice
analogue of the $Q$~transformation such that the nilpotency (up to the lattice
gauge transformation) holds. Then adopting a lattice action of the structure of
eq.~(\ref{twoxfour}), $Q$-invariance can be preserved exactly in lattice theory.

\subsection{Lattice formulation}

We consider two-dimensional square lattice of the one-dimensional physical
extent~$L$,
\begin{equation}
   \Lambda=\left\{x\in a\mathbb{Z}^2\mid 0\leq x_\mu<L\right\},
\label{twoxfive}
\end{equation}
where $a$~denotes the lattice spacing. We define also the one-dimensional
extent in a lattice unit $N=L/a$. All fields except the gauge potentials are
put on sites and, as is conventional in lattice gauge theory, the gauge field
is expressed by the compact link variables~$U(x,\mu)$. Periodic boundary
conditions on~$\Lambda$ are assumed on all fields.

As a lattice counterpart of the fermionic transformation~(\ref{twoxthree}), we
define ($\hat\mu$ implies a unit vector in the $\mu$-direction)
\begin{align}
   &QU(x,\mu)=i\psi_\mu(x)U(x,\mu),
\nonumber\\
   &Q\psi_\mu(x)=i\psi_\mu(x)\psi_\mu(x)
   -i\left(\phi(x)-U(x,\mu)\phi(x+a\hat\mu)U(x,\mu)^{-1}\right),
\nonumber\\
   &Q\phi(x)=0,
\nonumber\\
   &Q\chi(x)=H(x),\qquad QH(x)=[\phi(x),\chi(x)],
\nonumber\\
   &Q\overline\phi(x)=\eta(x),\qquad Q\eta(x)=[\phi(x),\overline\phi(x)].
\label{twoxsix}
\end{align}
It can be confirmed that $Q^2$ is in fact an infinitesimal lattice gauge
transformation with the parameter~$\phi(x)$. Thus the nilpotency $Q^2=0$ holds
on gauge invariant quantities. The lattice action is then defined by an
expression analogous to eq.~(\ref{twoxfour}):
\begin{equation}
   S=Qa^2\sum_{x\in\Lambda}\left(
   \mathcal{O}_1(x)+\mathcal{O}_2(x)+\mathcal{O}_3(x)
   +\frac{1}{a^4g^2}\tr\left\{\chi(x)H(x)\right\}\right),
\label{twoxseven}
\end{equation}
where
\begin{align}
   &\mathcal{O}_1(x)
   =\frac{1}{a^4g^2}
   \tr\left\{\frac{1}{4}\eta(x)[\phi(x),\overline\phi(x)]\right\},
\label{twoxeight}\\
   &\mathcal{O}_2(x)
   =\frac{1}{a^4g^2}
   \tr\left\{-i\chi(x)\hat\Phi(x)\right\},
\label{twoxnine}\\
   &\mathcal{O}_3(x)
   =\frac{1}{a^4g^2}
   \tr\left\{i\sum_{\mu=0}^1\psi_\mu(x)
   \left(\overline\phi(x)
   -U(x,\mu)\overline\phi(x+a\hat\mu)U(x,\mu)^{-1}\right)\right\}.
\label{twoxten}
\end{align}
In the above expression, $\hat\Phi(x)$ is a lattice analogue of the field
strength and is defined from the plaquette variables
\begin{equation}
    U(x,0,1)=U(x,0)U(x+a\hat0,1)U(x+a\hat1,0)^{-1}U(x,1)^{-1}
\label{twoxeleven}
\end{equation}
by
\begin{equation}
   \hat\Phi(x)
   =\frac{\Phi(x)}{1-\frac{1}{\epsilon^2}\left\|1-U(x,0,1)\right\|^2}
\label{twoxtwelve}
\end{equation}
with
\begin{equation}
   \Phi(x)=-i\left[U(x,0,1)-U(x,0,1)^{-1}\right].
\label{twoxthirteen}
\end{equation}
Finally, the matrix norm in the above expression is defined by
\begin{equation}
   \|A\|=\left[\tr\left\{AA^\dagger\right\}\right]^{1/2}
\label{twoxfourteen}
\end{equation}
and the constant~$\epsilon$ is chosen as (for $N_c=2$)
\begin{equation}
   0<\epsilon<2\sqrt{2}.
\label{twoxfifteen}
\end{equation}
(The meaning of the denominator of eq.~(\ref{twoxtwelve}) will be explained
shortly.) From the $Q$-exact form~(\ref{twoxseven}) and the nilpotency of $Q$,
the lattice action is manifestly invariant under the
$Q$-transformation~(\ref{twoxsix}).\footnote{Another important property of
the present lattice formulation is a manifestly preserved global
$\U(1)_R$ symmetry~\cite{Sugino:2003yb,Sugino:2004qd,Sugino:2004qdf}.}

After the operation of~$Q$, the lattice action becomes
\begin{equation}
   S=a^2\sum_{x\in\Lambda}\left(
   \sum_{i=1}^3\mathcal{L}_{\text{B}i}(x)+\sum_{i=1}^6\mathcal{L}_{\text{F}i}(x)
   +\frac{1}{a^4g^2}\tr\left\{H(x)-\frac{1}{2}i\hat\Phi_{\text{TL}}(x)\right\}^2
   \right),
\label{twoxsixteen}
\end{equation}
where we have noted that only the \emph{traceless part\/} of $\hat\Phi(x)$,
\begin{equation}
   \hat\Phi_{\text{TL}}(x)=\hat\Phi(x)
   -\frac{1}{N_c}\tr\left\{\hat\Phi(x)\right\}\mathbbm{1},
\label{twoxseventeen}
\end{equation}
appears in the action, because the auxiliary field~$H(x)$ is
traceless~\cite{Sugino:2004qdf}. Each term of the action density is given by
\begin{align}
   &\mathcal{L}_{\text{B}1}(x)=\frac{1}{a^4g^2}
   \tr\left\{\frac{1}{4}[\phi(x),\overline\phi(x)]^2\right\},
\label{twoxeighteen}\\
   &\mathcal{L}_{\text{B}2}(x)=\frac{1}{a^4g^2}
   \tr\left\{\frac{1}{4}\hat\Phi_{\text{TL}}(x)^2\right\},
\label{twoxnineteen}\\
   &\mathcal{L}_{\text{B}3}(x)=\frac{1}{a^4g^2}
   \tr\Biggl\{\sum_{\mu=0}^1
   \left(\phi(x)-U(x,\mu)\phi(x+a\hat\mu)U(x,\mu)^{-1}\right)
\nonumber\\
   &\qquad\qquad\qquad\qquad\qquad{}
   \times
   \left(\overline\phi(x)-U(x,\mu)\overline\phi(x+a\hat\mu)U(x,\mu)^{-1}
   \right)\Biggr\},
\label{twoxtwenty}
\end{align}
and
\begin{align}
   &\mathcal{L}_{\text{F}1}(x)=\frac{1}{a^4g^2}
   \tr\left\{-\frac{1}{4}\eta(x)[\phi(x),\eta(x)]\right\},
\label{twoxtwentyone}\\
   &\mathcal{L}_{\text{F}2}(x)=\frac{1}{a^4g^2}
   \tr\left\{-\chi(x)[\phi(x),\chi(x)]\right\},
\label{twoxtwentytwo}\\
   &\mathcal{L}_{\text{F}3}(x)=\frac{1}{a^4g^2}
   \tr\left\{-\psi_0(x)\psi_0(x)
   \left(\overline\phi(x)+U(x,0)\overline\phi(x+a\hat0)U(x,0)^{-1}
   \right)\right\},
\label{twoxtwentythree}\\
   &\mathcal{L}_{\text{F}4}(x)=\frac{1}{a^4g^2}
   \tr\left\{-\psi_1(x)\psi_1(x)
   \left(\overline\phi(x)+U(x,1)\overline\phi(x+a\hat1)U(x,1)^{-1}
   \right)\right\},
\label{twoxtwentyfour}\\
   &\mathcal{L}_{\text{F}5}(x)=\frac{1}{a^4g^2}
   \tr\left\{i\chi(x)Q\hat\Phi(x)\right\},
\label{twoxtwentyfive}\\
   &\mathcal{L}_{\text{F}6}(x)=\frac{1}{a^4g^2}
   \tr\left\{-i\sum_{\mu=0}^1\psi_\mu(x)
   \left(\eta(x)-U(x,\mu)\eta(x+a\hat\mu)U(x,\mu)^{-1}\right)\right\}.
\label{twoxtwentysix}
\end{align}
It is important to keep in mind that all lattice fields in the above
expressions are dimensionless. For comparison of correlation functions with the
continuum theory~(\ref{twoxtwo}), we have to rescale all lattice fields (and
$Q$) by appropriate factors of $1/a$ according to their mass dimension.

With the lattice action~(\ref{twoxsixteen}), the expectation value of an
operator~$\mathcal{O}$ is defined as usual
\begin{equation}
   \llangle\mathcal{O}\rrangle
   =\frac{\displaystyle\int\rmd\mu\,\mathcal{O}\,e^{-S}}
   {\displaystyle\int\rmd\mu\,e^{-S}},
\label{twoxtwentyseven}
\end{equation}
where the integration measure is defined by (writing
$\phi(x)=X_2(x)+iX_3(x)$ and $\overline\phi(x)=X_2(x)-iX_3(x)$)
\begin{equation}
   \rmd\mu=\prod_{x\in\Lambda}\left(\prod_{\mu=0}^1\rmd U(x,\mu)\right)
   \prod_{a=1}^{N_c^2-1}\rmd X_2^a(x)\,\rmd X_3^a(x)\,\rmd H^a(x)
   \left(\prod_{\mu=0}^1\rmd\psi_\mu^a(x)\right)
   \rmd\chi^a(x)\,\rmd\eta^a(x)
\label{twoxtwentyeight}
\end{equation}
in terms of color components of fields,
$\varphi(x)=-i\sum_{a=1}^{N_c^2-1}\varphi^a(x)T^a$, where $T^a$ are
anti-hermitian generators of $\SU(N_c)$ (normalized as
$\tr\{T^aT^b\}=-(1/2)\delta_{ab}$). $\rmd U(x,\mu)$ is the standard Haar
measure. Note that the integration over the auxiliary field~$H(x)$ is gaussian
and can be done readily. The invariance of this measure under the
$Q$-transformation is noted in the last reference of
ref.~\cite{Sugino:2003yb}.

The denominator of eq.~(\ref{twoxtwelve}) needs an explanation. Without that
factor, the lattice action for the gauge field is the ``double-winding
plaquette type''~\cite{Elitzur:1982vh} and the action possesses many degenerate
minima which have no continuum counterpart. Due to the denominator of
eq.~(\ref{twoxtwelve}), the action~(\ref{twoxsixteen}) diverges as
$\left\|1-U(x,0,1)\right\|\to\epsilon$ at a certain site~$x$. Precisely
speaking, the above construction of the action is applied only for
configurations with
\begin{equation}
     \left\|1-U(x,0,1)\right\|<\epsilon,\qquad\hbox{for $\forall x\in\Lambda$},
\label{twoxtwentynine}
\end{equation}
and, otherwise, i.e., if there exists $x\in\Lambda$ such that
$\left\|1-U(x,0,1)\right\|\geq\epsilon$, we set
\begin{equation}
   S=+\infty.
\label{twoxthirty}
\end{equation}
In this way, the domain of functional integral~(\ref{twoxtwentyseven}) is
effectively restricted to the space specified by the
condition~(\ref{twoxtwentynine})\footnote{This is the so-called admissibility
condition considered in a different context~\cite{Luscher:1999du}.} and, by
setting the parameter~$\epsilon$ in the range~(\ref{twoxfifteen}), it can be
shown that the unique physical minimum of the action (up to gauge
transformations) is singled out. This procedure to solve the problem of
degenerate minima does not break the $Q$-symmetry. See
ref.~\cite{Sugino:2004qdf} for careful discussion on these points.

With the above construction, one fermionic symmetry~$Q$ is manifestly
preserved on the lattice. The price to pay is that the pfaffian, resulting from
the integration over fermionic fields, is generally complex\footnote{In the
target continuum theory, the corresponding pfaffian is real and positive
semi-definite.} and this could be disadvantage in Monte Carlo simulation. We
will see below that, however, this point appears to be not problematic, at
least with the parameters in our numerical study.

\subsection{Continuum limit}
In the present two-dimensional super-renormalizable system, all dimension-ful
quantities can be measured by taking the gauge coupling constant~$g$, which has
the mass dimension~1, as a unit (in this sense, $g$ is analogous to the
$\Lambda$-parameter in QCD).\footnote{Recall also that, in the present system,
there is no non-trivial coupling constant renormalization nor wave function
renormalization. Only mass terms of bosonic fields may be renormalized
(ignoring gauge symmetry and supersymmetry).} The continuum limit is defined by
the limit~$a\to0$, while $g$ and $L$, the physical extent of the
two-dimensional space, are kept fixed. In
refs.~\cite{Sugino:2003yb,Sugino:2004qd}, the restoration of \emph{full\/} set
of supersymmetry in this continuum limit was argued on the basis of the loop
expansion and power counting.\footnote{Strictly speaking, this argument as it
stands holds for the limit $a\to0$ with the fixed number of lattice points
$N=L/a$ (thus the physical lattice size goes to zero $L=aN\to0$). The argument,
however, can slightly be modified to show a restoration of supersymmetry in the
present ($L$ fixed) continuum limit, to all orders of the loop expansion.}
More precisely, this argument shows that the 1PI effective action for
\emph{elementary fields\/} is supersymmetric in the continuum limit. Note that
the argument of refs.~\cite{Sugino:2003yb,Sugino:2004qd} says nothing about
possible supersymmetry breaking in correlation functions that contain
\emph{composite fields}.

Now, in numerical study, it is convenient to define the dimensionless gauge
coupling constant by
\begin{equation}
   \frac{\beta}{2N_c}=\frac{1}{a^2g^2},
\label{twoxthirtyone}
\end{equation}
that is simply the over-all common coefficient of the lattice
action~(\ref{twoxsixteen}). Clearly, $\beta$ goes infinity in the continuum
limit. In terms of~$\beta$, the lattice spacing in a unit of the gauge coupling
constant~$g$ is given by
\begin{equation}
   a=\sqrt{\frac{2N_c}{\beta}}\frac{1}{g},
\label{twoxthirtytwo}
\end{equation}
and, correspondingly, the one-dimensional physical extent of the lattice is
\begin{equation}
   L=aN=\sqrt{\frac{2N_c}{\beta}}N\frac{1}{g},
\label{twoxthirtythree}
\end{equation}
where $N$ is the one-dimensional size in a lattice unit.

As already noted, all fields on the lattice must be rescaled by appropriate
factors of~$1/a$, for comparison with the continuum theory~(\ref{twoxtwo}). All
bosonic fields in the continuum theory (except the auxiliary field), which have
the mass dimension~1, are related to the lattice fields by
\begin{equation}
   \varphi_{\text{continuum}}(x)=\frac{1}{a}\varphi(x)
   =\sqrt{\frac{\beta}{2N_c}}g\varphi(x)
\label{twoxthirtyfour}
\end{equation}
and the correlation functions are measured in a unit of~$g$. Similarly,
fermionic fields are related as
\begin{equation}
   \psi_{\text{continuum}}(x)=\frac{1}{a^{3/2}}\psi(x)
   =\left(\frac{\beta}{2N_c}\right)^{3/2}g^{3/2}\psi(x).
\label{twoxthirtyfive}
\end{equation}
Note that, in the continuum limit, these rescalings \emph{amplify\/} the
correlation functions on the lattice.

\section{Monte Carlo study}
\subsection{Algorithm, simulation code and statistics}
In supersymmetric theories, the quantum effect of fermions is vital and the
quenched approximation is almost meaningless. Even in the present
two-dimensional system, a treatment of dynamical fermions can be non-trivial
and costly. The 2d $\mathcal{N}=(2,2)$ SYM can be obtained by dimensional
reduction of the 4d $\mathcal{N}=1$ SYM in which the fermion field is a
\emph{Majorana\/} spinor instead of Dirac. Thus the pfaffian of the Dirac
operator, instead of the determinant, naturally appears. In a sense, we have to
treat an $N_f=1/2$ system. To compute the pseudo-fermion force in the hybrid
Monte Carlo algorithm, one then has to implement the fourth-root of
$D^\dagger D$, where $D$ is a lattice Dirac operator. Moreover, this fermion
must be massless (at least in the continuum limit). Thus the numerical
simulation of the 4d SYM is quite demanding.

In two dimensions, on the other hand, it should be relatively easy to
accumulate high statistics compared to four dimensions. Taking these things
into consideration, here we adopt a (somewhat brute force) re-weighting
method.\footnote{Hidenori Fukaya suggested this method to me. For application
in two dimensions, see, for example, ref.~\cite{Durr:2003xs}.} That is, we
prepare configurations with the statistical weight~$e^{-S_{\text{B}}}$, where
$S_{\text{B}}$ is the lattice action~(\ref{twoxsixteen}) with all fermion fields
are removed. This is a quenched ensemble. Writing the expectation value in this
purely bosonic system by\footnote{When the operator~$\mathcal{O}$ contains
fermionic fields, they are contracted by fermion propagators in the presence of
bosonic fields.}
\begin{equation}
   \left\langle\mathcal{O}\right\rangle
   =\frac{\displaystyle\int\rmd\mu_{\text{B}}\,\mathcal{O}\,e^{-S_{\text{B}}}}
   {\displaystyle\int\rmd\mu_{\text{B}}\,e^{-S_{\text{B}}}},
\label{threexone}
\end{equation}
the true expectation value is evaluated by re-weighting configurations by the
factor of pfaffian\footnote{Note that, in this method,
$\llangle\mathcal{O}\rrangle$ is evaluated by a \emph{ratio\/} of two averages
over an ensemble. This means that $\llangle\mathcal{O}\rrangle$ is not
the primary quantity~\cite{Montvay:1994cy} and care is needed to estimate the
statistical error in~$\llangle\mathcal{O}\rrangle$. We used the jackknife
analysis to estimate the average and the statistical error for
$\llangle\mathcal{O}\rrangle$. I would like to thank Issaku Kanamori for
clarifying discussion on this point.}
\begin{equation}
   \llangle\mathcal{O}\rrangle
   =\frac{\displaystyle\left\langle\mathcal{O}\Pf\{D\}\right\rangle}
   {\displaystyle\left\langle\Pf\{D\}\right\rangle},
\label{threextwo}
\end{equation}
where $D$ is the lattice Dirac operator appeared in the
action~(\ref{twoxsixteen}). Mathematically, this definition is equivalent to
the original one~(\ref{twoxtwentyseven}). Practically, however, we have only a
limited number of configurations and there may exist the overlap problem.
That is, distribution of configurations favored by the the quenched
weight~$e^{-S_{\text{B}}}$ may not have a sufficient overlap with that of
configurations really important in the original system. So we need many
configurations to reproduce the true expectation values in the original
un-quenched system.

We developed a C++ code of the hybrid Monte Carlo simulation with the
action~$S_{\text{B}}$ by using a library due to Massimo Di Pierro, the
FermiQCD/MDP~\cite{DiPierro:2000bd}. For each configuration, we compute the
inverse (i.e., the fermion propagator) and the determinant of the lattice Dirac
operator~$D$ by using the LU decomposition. \emph{We do not introduce any
(supersymmetry breaking) mass terms of fermions and bosons}.

\TABULAR[t]{|c|c|c|c|c|c|}{\hline 
$N$ & 8 & 7 & 6 & 5 & 4 \\
\hline
$\beta$ & $16.0$ & $12.25$ & $9.0$ & $6.25$ & $4.0$ \\
\hline
number of configs. & 1,000 & 10,000 & 10,000 & 10,000 & 10,000 \\
\hline
$ag$ & $0.5$ & $0.571428$ & $0.666666$ & $0.8$ & $1.0$ \\
\hline}{Parameters in our Monte Carlo study. This sequence corresponds to
the fixed physical lattice size~$Lg=4.0$.\label{table1}}

We carried out simulations with the parameters in table~\ref{table1}. The
sequence, according to eqs.~(\ref{twoxthirtytwo}) and~(\ref{twoxthirtythree}),
corresponds to a fixed physical lattice size $Lg=4.0$ and the lattice spacings
$ag=1.0$, $0.8$, $0.666$, $0.571$ and~$0.5$, respectively. For each value of
$\beta$, we stored 1,000--10,000 independent configurations extracted from
$10^6$ trajectories of the molecular dynamics.%
\ifthenelse{\value{private} = 100}{%
\footnote{Time step of the leapfrog was
fixed to~0.1 and the number of steps in one trajectory was taken to be~10. The
Metropolis acceptance was 63\%--96\% depending on the parameter.}
}{}
The constant $\epsilon$ in eq.~(\ref{twoxtwelve}) is kept fixed at
$\epsilon=2.6$.\footnote{We observed a tendency such that the
autocorrelation time becomes shorter for smaller $\epsilon$. Thus
small~$\epsilon$ would be favorable from a viewpoint to accumulate a large
number of configurations. On the other hand, it appears that smaller~$\epsilon$
implies smaller fluctuation of distribution of configurations and might be
disadvantageous from a viewpoint of the overlap problem. We did not
systematically investigate this problem of an optimal~$\epsilon$. Our present
$\epsilon$ is rather large in view of eq.~(\ref{twoxfifteen}).}

Expressing the determinant of the Dirac operator in the form
\begin{equation}
   \det\{D\}=re^{i\theta},\qquad-\pi<\theta\leq\pi, 
\label{threexthree}
\end{equation}
(generally the determinant is complex due to lattice artifacts in the present
formulation) we evaluate the pfaffian by
\begin{equation}
   \Pf\{D\}=\sqrt{r}e^{i\theta/2},
\label{threexfour}
\end{equation}
because $(\Pf\{D\})^2=\det\{D\}$. This prescription, however, may give a
wrong \emph{sign\/} for the pfaffian. For example, if
$\Pf\{D\}=\sqrt{r}e^{(2/3)\pi i}$, we have $\theta=-2\pi/3$ and the
prescription~(\ref{threexfour}) gives
$\Pf\{D\}=\sqrt{r}e^{-(1/3)\pi i}=-\sqrt{r}e^{(2/3)\pi i}$ which is wrong in sign.
The prescription~(\ref{threexfour}) gives the correct sign of the pfaffian,
provided that $-\pi/2<\Arg(\Pf\{D\})\leq\pi/2$ (and otherwise the prescription
gives a wrong sign). Although this is expected to be the case for
large~$\beta$ (i.e., when close to the continuum), to determine the true sign
of the pfaffian, we have to compute the pfaffian itself in some direct way.
This is quite time-consuming\footnote{It can be seen that the algorithm for the
pfaffian (appearing, for example, in ref.~\cite{Campos:1999du}) is an
$O(n^4)$-process for a $2n\times2n$ matrix, while the LU decomposition has an
$O(n^3)$-process algorithm.}
and we do not adopt this method in this paper. Instead, to have an idea how the
prescription~(\ref{threexfour}) works in practice, we measured the distribution
of the pfaffians over a subset of our ensemble in table~\ref{table1}.
\FIGURE[t]{\centerline{\epsfig{file=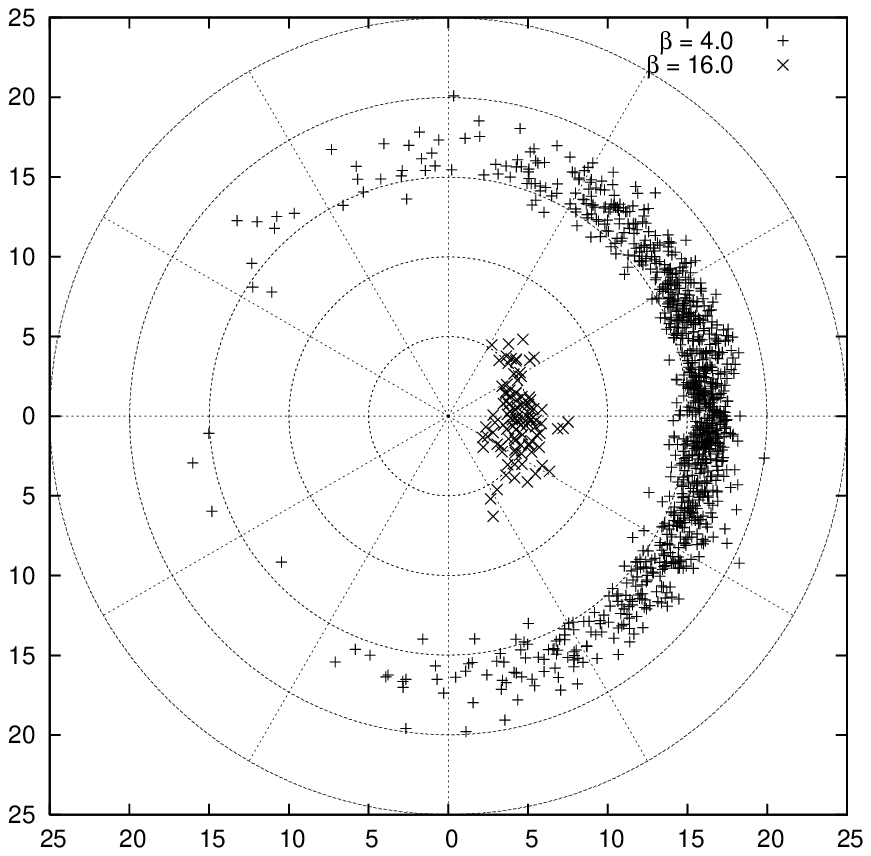,width=0.8\textwidth}}%
\caption{The distribution of pfaffians in a subset of the quenched ensemble
used in our simulation. The phase, $\Arg(\Pf\{D\})$, and the modulus in
logarithm, $\log_{10}(10^{16}|\Pf\{D\}|)$, are plotted in the polar coordinate.
The number of samples is 1,000 and 100 for $\beta=4.0$ and $\beta=16.0$,
respectively.\label{figure1}}}
The behavior in the figure~\ref{figure1} clearly accords with our expectation.
For large $\beta$ (i.e., close to the continuum), the distribution gathers
around the positive side of the real axis and the condition
$-\pi/2<\Arg(\Pf\{D\})\leq\pi/2$ is fulfilled. Even for the smallest~$\beta$
in our simulation, $\beta=4.0$, the distribution is significantly biased on the
side of the positive real axis. Thus the systematic error introduced by the
wrong-sign determination due to the prescription~(\ref{threexfour}) would be
negligible compared to the statistical error.

We consider also the quenched approximation, i.e.,
\begin{equation}
   \llangle\mathcal{O}\rrangle_{\text{quenched}}
   =\frac{\displaystyle\left\langle\mathcal{O}\right\rangle}
   {\displaystyle\left\langle1\right\rangle}.
\label{threexsix}
\end{equation}
This provides a useful standard with which one can observe the extent of the
quantum effect of dynamical fermions.

\subsection{One-point supersymmetric WT identities}
First, we consider supersymmetric one-point WT identities implied by the exact
$Q$-invariance of the lattice action. These are of the form
$\llangle Q(\text{something})\rrangle$ and identically vanish because of
$Q$-invariance of the action and the integration measure. These should hold for
any lattice parameter, if the integration (especially that over fermionic
fields) is properly performed. Thus, from their validity in numerical
simulation, we can confirm the correctness of our code/algorithm. In
particular, we can observe whether the re-weighting method works or not.

Since the lattice action~(\ref{twoxseven}) is $Q$-exact, we have
$\llangle S\rrangle=0$, or, in terms of the action density,
\begin{equation}
   \sum_{i=1}^3\llangle\mathcal{L}_{\text{B}i}(x)\rrangle
   +\sum_{i=1}^6\llangle\mathcal{L}_{\text{F}i}(x)\rrangle
   +\frac{1}{a^4g^2}
   \llangle\tr\left\{H(x)-\frac{1}{2}i\hat\Phi_{\text{TL}}(x)\right\}^2
   \rrangle=0.
\label{threexseven}
\end{equation}
One may further simplify this relation. The second term is the expectation
value of the action density of the fermionic fields. Since the action is
bi-linear in fermionic fields, we have
$\sum_{i=1}^6\llangle\mathcal{L}_{\text{F}i}(x)\rrangle=-2(N_c^2-1)a^{-2}$ (the
coefficient 2 is $(1/2)\times 4$, where $1/2$ reflects the Majorana nature
of the system and 4 is the number of fermion species). Similarly, the auxiliary
field~$H(x)$ can be integrated out and the last term becomes
$(1/2)(N_c^2-1)a^{-2}$ after integration. Thus
\begin{equation}
   \sum_{i=1}^3\llangle\mathcal{L}_{\text{B}i}(x)\rrangle
   -\frac{3}{2}(N_c^2-1)\frac{1}{a^2}=0.
\label{threexeight}
\end{equation}
In figure~\ref{figure2}, we plotted the left-hand side of this relation
(in a unit of~$g^2$) as a function of the lattice spacing~$ag$.
\FIGURE[t]{\centerline{\epsfig{file=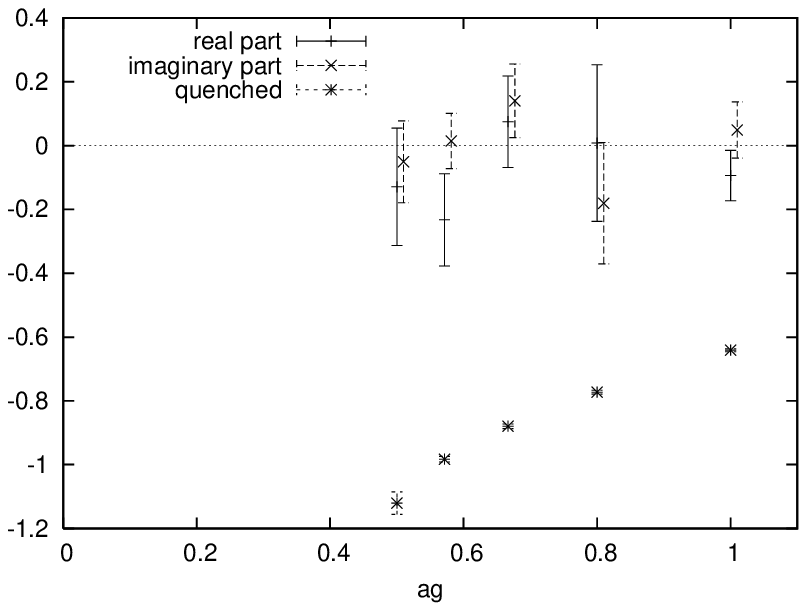,width=0.8\textwidth}}%
\caption{Expectation values of
$\sum_{i=1}^3\mathcal{L}_{\text{B}i}(x)-(3/2)(N_c^2-1)a^{-2}$.
\label{figure2}}}
The real part is consistent with the expected identity~(\ref{threexeight})
within $1\sigma$ for all values of $ag$, except $ag=0.571$ that is $1.5\sigma$
away. This agreement strongly indicates the correctness of our
code/algorithm. The average of the imaginary part is consistent with zero, as
it should be, although its fluctuation is comparable to that of the real part.

What is intriguing with figure~\ref{figure2} is that one can see clear
distinction between the re-weighted average~(\ref{threextwo}) and the quenched
average~(\ref{threexsix}). This illustrates that the re-weighting method works
very well and the effect of dynamical fermions is properly included (at least
for the present quantity). A perturbative argument shows that each term of the
action density behaves as
\begin{equation}
   \llangle\mathcal{L}_{\text{B}2}(x)\rrangle
   \sim\frac{1}{2}(N_c^2-1)\frac{1}{a^2},\qquad
   \llangle\mathcal{L}_{\text{B}3}(x)\rrangle\sim(N_c^2-1)\frac{1}{a^2},
\label{threexnine}
\end{equation}
in $a\to0$ because of one-loop diagrams and
$\llangle\mathcal{L}_{\text{B}1}(x)\rrangle$ starts with a two-loop diagram
which behaves as~$\sim(\ln(a/L))^2g^2$. The leading $O(a^{-2})$ singularities
are thus cancelled out in the sum
$\sum_{i=1}^3\llangle\mathcal{L}_{\text{B}i}(x)\rrangle-(3/2)(N_c^2-1)a^{-2}$ and
this leaves a function of the form $f(a/L,Lg)g^2$. This function identically
vanishes if supersymmetry holds, but it is a non-trivial function in the
quenched approximation. What is shown in the figure~\ref{figure2} with
``quenched'' is this function.\footnote{The lattice perturbation theory is not
useful to evaluate this function even for $ag\to0$. In the loop expansion, we
may have terms of the form, say, $(\ln(a/L))^\ell(Lg)^{2(\ell-2)}g^2$ at
$\ell$-loop level. All this type of terms equally contribute to the function
$f$ in the present continuum limit, in which $Lg$ is fixed (recall that
$Lg=4.0$ in our simulation).}

The identity~(\ref{threexseven}) can be divided into several pieces, each of
which should hold separately. The first one is
\begin{equation}
   \llangle Q\mathcal{O}_1(x)\rrangle
   =\llangle\mathcal{L}_{\text{B}1}(x)\rrangle
   +\llangle\mathcal{L}_{\text{F}1}(x)\rrangle=0,
\label{threexten}
\end{equation}
and the left-hand side is plotted in figure~\ref{figure3}.
\FIGURE[t]{\centerline{\epsfig{file=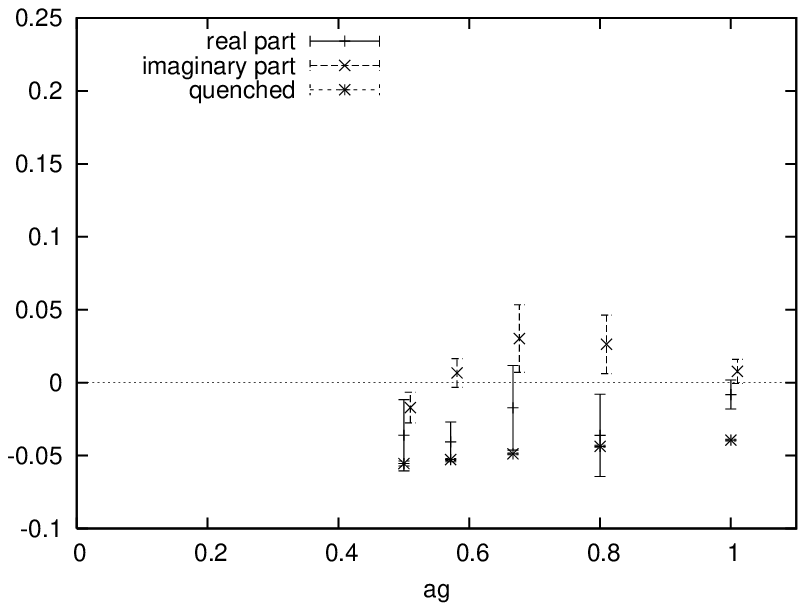,width=0.8\textwidth}}%
\caption{Expectation values of
$\mathcal{L}_{\text{B}1}(x)+\mathcal{L}_{\text{F}1}(x)$.
\label{figure3}}}
The relation is confirmed within $1.5\sigma$ expect the case~$ag=0.571$. Note
the difference in scale of vertical axis compared to figure~\ref{figure2}.
Although the results with a quenched ensemble are certainly inconsistent with
the supersymmetric relation~(\ref{threexten}), we do not see clear separation
between the re-weighted average and the quenched average. This seems to be
related to the fact that diagrams that contribute to
$\llangle\mathcal{L}_{\text{B}1}(x)\rrangle$ and
$\llangle\mathcal{L}_{\text{F}1}(x)\rrangle$ and contain virtual fermion loops
start with three loops, a rather higher order.

Another piece of eq.~(\ref{threexseven}) is
\begin{equation}
   \llangle Q\mathcal{O}_2(x)\rrangle
   =\frac{1}{a^4g^2}\llangle\tr\left\{-iH(x)\hat\Phi_{\text{TL}}(x)\right\}
   \rrangle
   +\llangle\mathcal{L}_{\text{F}5}(x)\rrangle=0.
\label{threexeleven}
\end{equation}
Under the gaussian integration, the auxiliary field can be replaced by
$H(x)=\frac{1}{2}i\hat\Phi_{\text{TL}}(x)$ and the above becomes
\begin{equation}
   2\llangle\mathcal{L}_{\text{B}2}(x)\rrangle
   +\llangle\mathcal{L}_{\text{F}5}(x)\rrangle=0.
\label{threextwelve}
\end{equation}
In figure~\ref{figure4}, the left-hand side of this relation is plotted.
\FIGURE[t]{\centerline{\epsfig{file=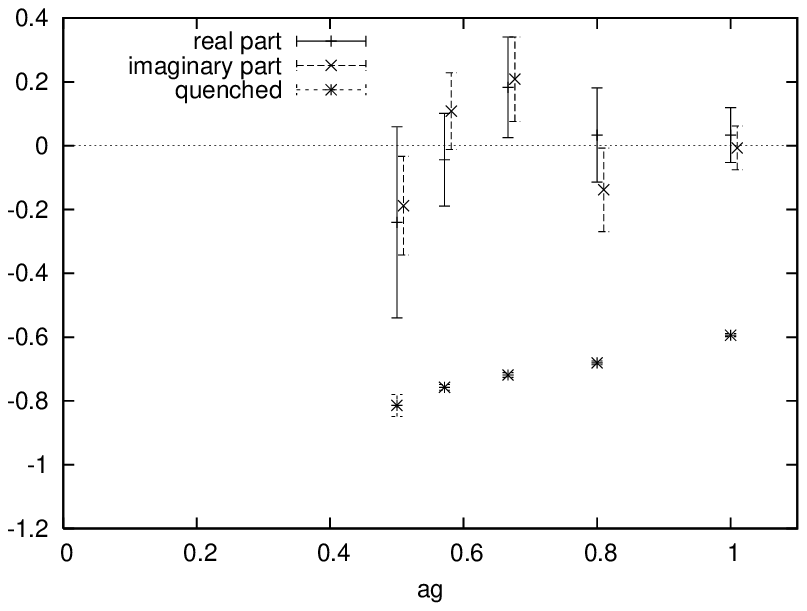,width=0.8\textwidth}}%
\caption{Expectation values of
$2\mathcal{L}_{\text{B}2}(x)+\mathcal{L}_{\text{F}5}(x)$.
\label{figure4}}}
The global feature is similar to that of figure~\ref{figure2} and the relation
is reproduced within almost $1\sigma$.

The situation is again similar with the last piece of the
relation~(\ref{threexseven}):
\begin{equation}
   \llangle Q\mathcal{O}_3(x)\rrangle
   =\llangle\mathcal{L}_{\text{B}3}(x)\rrangle
   +\llangle\mathcal{L}_{\text{F}3}(x)\rrangle
   +\llangle\mathcal{L}_{\text{F}4}(x)\rrangle
   +\llangle\mathcal{L}_{\text{F}6}(x)\rrangle=0,
\label{threexthirteen}
\end{equation}
whose left-hand side is plotted in figure~\ref{figure5}.
\FIGURE[t]{\centerline{\epsfig{file=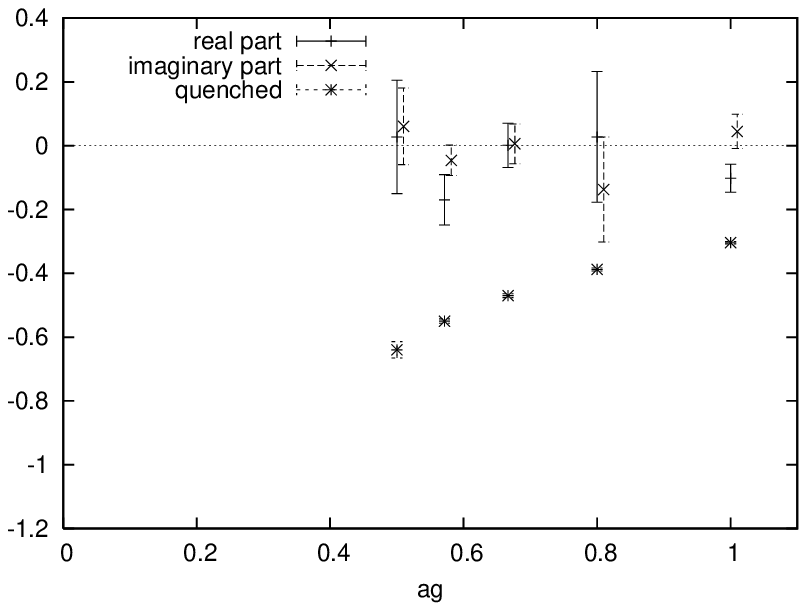,width=0.8\textwidth}}%
\caption{Expectation values of
$\mathcal{L}_{\text{B}3}(x)+\mathcal{L}_{\text{F}3}(x)
+\mathcal{L}_{\text{F}4}(x)+\mathcal{L}_{\text{F}6}(x)$.
\label{figure5}}}

So far, we have observed WT identities implied by the exact $Q$-symmetry of the
lattice action. The continuum theory~(\ref{twoxtwo}), on the other hand, is
invariant under also other fermionic transformations, $Q_{01}$, $Q_0$ and~$Q_1$.
In the lattice framework, the invariance under these transformations is
expected to be restored only in the continuum limit. The fermionic
transformation $Q_{01}$ is given by
\begin{align}
   &Q_{01}A_\mu=-\epsilon_{\mu\nu}\psi_\mu,
   &&Q_{01}\psi_\mu=i\epsilon_{\mu\nu}D_\nu\phi,
\nonumber\\
   &Q_{01}\phi=0,&&
\nonumber\\
   &Q_{01}\eta=2H,&&Q_{01}H=\frac{1}{2}[\phi,\eta],
\nonumber\\
   &Q_{01}\overline\phi=-2\chi,&&Q_{01}\chi=-\frac{1}{2}[\phi,\overline\phi],
\label{threexfourteen}
\end{align}
that can be obtained by following substitutions in the
$Q$-transformation~(\ref{twoxthree})
\begin{equation}
   \frac{1}{2}\eta\to-\chi,\qquad\chi\to\frac{1}{2}\eta,\qquad
   \psi_\mu\to-\epsilon_{\mu\nu}\psi_\nu,
\label{threexfifteen}
\end{equation}
where $\epsilon_{01}=-\epsilon_{10}=1$. Since the action~(\ref{twoxtwo}) is
invariant under these substitutions, the invariance of the continuum action
under eq.~(\ref{threexfourteen}) is obvious. Associated with this
$Q_{01}$-invariance, in the supersymmetric continuum theory, we have
\begin{align}
   &\llangle Q_{01}\frac{1}{g^2}
   \tr\left\{-\frac{1}{2}\chi[\phi,\overline\phi]\right\}
   \rrangle_{\text{continuum}}
\nonumber\\
   &=\frac{1}{g^2}\llangle
   \tr\left\{\frac{1}{4}[\phi,\overline\phi]^2\right\}
   \rrangle_{\text{continuum}}
   +\frac{1}{g^2}\llangle
   \tr\left\{-\chi[\phi,\chi]\right\}
   \rrangle_{\text{continuum}}=0.
\label{threexsixteen}
\end{align}
Thus, corresponding to this relation, one might expect
\begin{equation}
   \llangle\mathcal{L}_{\text{B}1}(x)\rrangle
   +\llangle\mathcal{L}_{\text{F}2}(x)\rrangle\to0
\label{threexseventeen}
\end{equation}
holds in the continuum limit~$a\to0$.

In figure~\ref{figure6}, we plotted the left-hand side of
eq.~(\ref{threexseventeen}).
\FIGURE[t]{\centerline{\epsfig{file=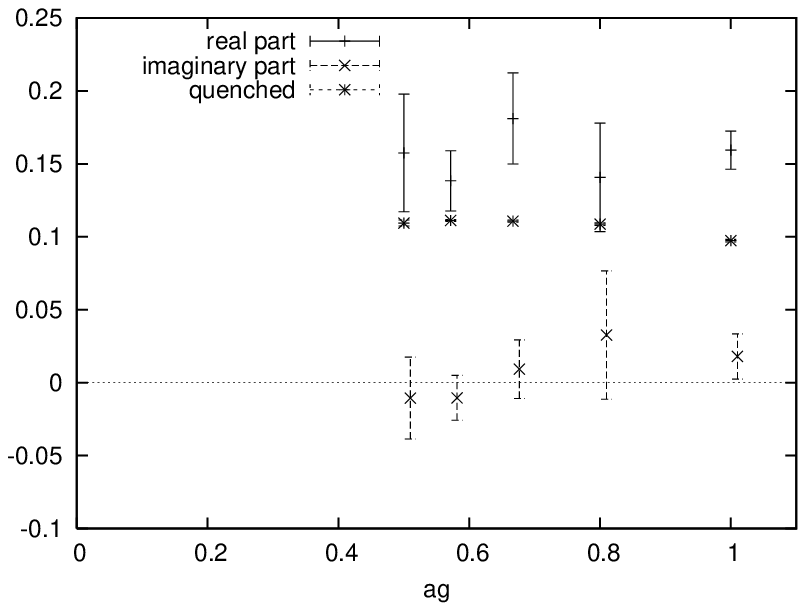,width=0.8\textwidth}}%
\caption{Expectation values of
$\mathcal{L}_{\text{B}1}(x)+\mathcal{L}_{\text{F}2}(x)$.
\label{figure6}}}
It appears that the average approaches a non-zero number around $0.15$, instead
of zero (the imaginary part is consistent with zero, as it should be).
This does \emph{not\/} contradict with the supersymmetry restoration. As
already noted, the argument~\cite{Sugino:2003yb} for a restoration of
supersymmetry in the continuum limit is not applied to correlation functions
containing composite operators. In particular, there is no general guarantee
that the bare WT identity~(\ref{threexseventeen}) holds in the continuum limit.

We note that if supersymmetry in the 1PI effective action is restored in the
continuum limit, it is UV finite, that is, all 1PI diagrams with external lines
of elementary fields are UV finite. Power counting (taking gauge invariance
into account) shows that only scalar mass terms suffer from superficial UV
divergence. Scalar mass terms are, however, inconsistent with supersymmetry. So
if the 1PI effective action is supersymmetric, it is UV finite.\footnote{The
contrary is not true. The UV finiteness of the effective action does not imply
supersymmetry, as finite scalar mass terms are allowed for the former.} On the
other hand, composite
operators~$\mathcal{L}_{\text{B}1}(x)$ and~$\mathcal{L}_{\text{F}2}(x)$ induce
logarithmic UV divergence at two-loop level. If supersymmetry of the 1PI
effective action is restored, this two-loop level divergence, caused by the
presence of composite operators, is the only source of UV divergence in
$\llangle\mathcal{L}_{\text{B}1}(x)\rrangle$ and
$\llangle\mathcal{L}_{\text{F}2}(x)\rrangle$. Moreover, that
remaining two-loop level divergence is cancelled out in the sum
$\llangle\mathcal{L}_{\text{B}1}(x)\rrangle
+\llangle\mathcal{L}_{\text{F}2}(x)\rrangle$.
This argument shows that, if supersymmetry in the 1PI effective action
restores, the dependence of $\llangle\mathcal{L}_{\text{B}1}(x)\rrangle
+\llangle\mathcal{L}_{\text{F}2}(x)\rrangle$ on~$ag$ decreases as $ag\to0$,
i.e., it approaches a constant (but not necessarily zero). The behavior in
figure~\ref{figure6} is consistent with this picture based on a restoration of
supersymmetry.

What is not completely clear to us is that even the quenched average seems to
have the same behavior. Actually, within almost $1\sigma$ the re-weighted
average and the quenched average are degenerate. So, although
figure~\ref{figure6} is consistent with a scenario of a supersymmetry
restoration, we cannot conclude the restoration of supersymmetry from the above
result.

The continuum action is invariant under also
\begin{align}
   &Q_0A_0=\frac{1}{2}\eta,&&Q_0\eta=-2iD_0\overline\phi,
\nonumber\\
   &Q_0A_1=-\chi,&&Q_0\chi=iD_1\overline\phi,
\nonumber\\
   &Q_0\overline\phi=0,&&
\nonumber\\
   &Q_0\psi_1=-H,&&Q_0H=[\overline\phi,\psi_1],
\nonumber\\
   &Q_0\phi=-2\psi_0,&&Q_0\psi_0=\frac{1}{2}[\overline\phi,\phi],
\label{threexeighteen}
\end{align}
that can be obtained by the substitutions in eq.~(\ref{twoxthree})
\begin{equation}
   \frac{1}{2}\eta\to\psi_0,\qquad \chi\to-\psi_1,\qquad
   \psi_0\to\frac{1}{2}\eta,\qquad\psi_1\to-\chi,\qquad
   \phi\to-\overline\phi,\qquad\overline\phi\to-\phi.
\label{threexnineteen}
\end{equation}
Corresponding to this symmetry, we have
\begin{align}
   &\llangle Q_0\frac{1}{g^2}
   \tr\left\{-\frac{1}{2}\psi_0[\phi,\overline\phi]\right\}
   \rrangle_{\text{continuum}}
\nonumber\\
   &=\frac{1}{g^2}\llangle
   \tr\left\{\frac{1}{4}[\phi,\overline\phi]^2\right\}
   \rrangle_{\text{continuum}}
   +\frac{1}{g^2}\llangle
   \tr\left\{-\psi_0[\psi_0,\overline\phi]\right\}
   \rrangle_{\text{continuum}}=0
\label{threextwenty}
\end{align}
and one might expect
\begin{equation}
   \llangle\mathcal{L}_{\text{B}1}(x)\rrangle
   +\llangle\mathcal{L}_{\text{F}3}(x)\rrangle\to0,
\label{threextwentyone}
\end{equation}
in the continuum limit $a\to0$. The result (figure~\ref{figure7}) is similar to
the previous one.
\FIGURE[t]{\centerline{\epsfig{file=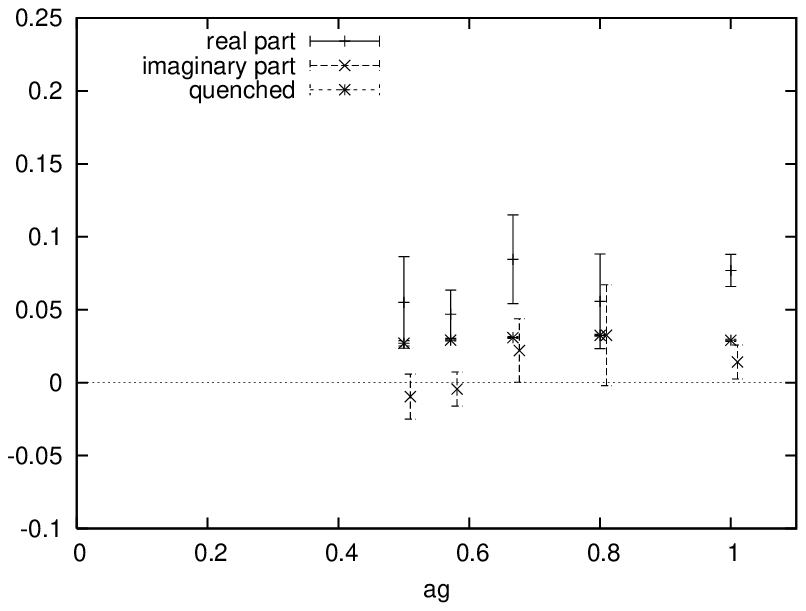,width=0.8\textwidth}}%
\caption{Expectation values of
$\mathcal{L}_{\text{B}1}(x)+\mathcal{L}_{\text{F}3}(x)$.
\label{figure7}}}
The average seems to approach a non-zero number around $0.05$ and we may repeat
the above argument.

Another fermionic symmetry~$Q_1$ is obtained by further exchange
$\psi_0\leftrightarrow\psi_1$ in eq.~(\ref{threexnineteen}). Corresponding
to this, one might expect
\begin{equation}
   \llangle\mathcal{L}_{\text{B}1}(x)\rrangle
   +\llangle\mathcal{L}_{\text{F}4}(x)\rrangle\to0.
\label{threextwentytwo}
\end{equation}
The result of numerical study is plotted in figure~\ref{figure8}.
\FIGURE[t]{\centerline{\epsfig{file=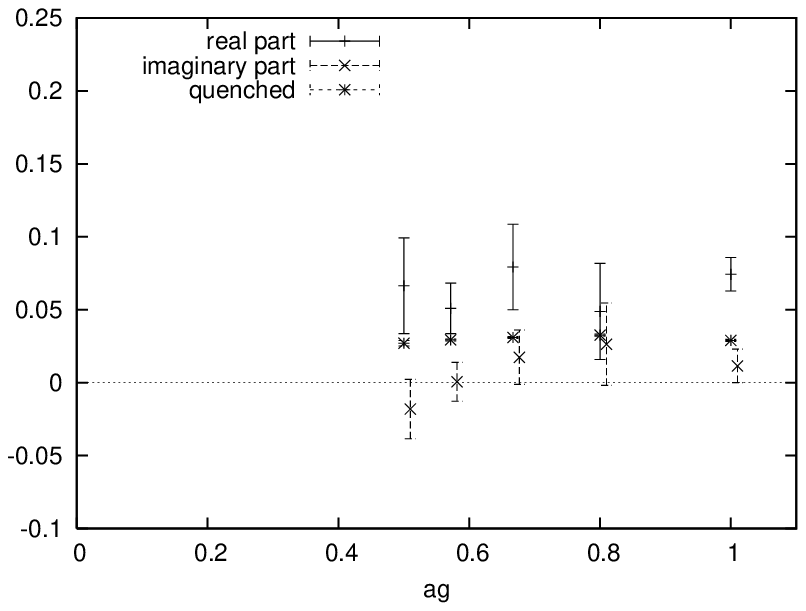,width=0.8\textwidth}}%
\caption{Expectation values of
$\mathcal{L}_{\text{B}1}(x)+\mathcal{L}_{\text{F}4}(x)$.
\label{figure8}}}
The result is very similar to that of figure~\ref{figure7}.

\subsection{Expectation value of scalar bi-linear operators}
To illustrate possible use of lattice simulation of the present kind, in this
section, we consider expectation values of gauge-invariant bi-linear operators
of scalar fields, $a^{-2}\tr\{\phi(x)\overline\phi(x)\}$
and~$a^{-2}\tr\{\phi(x)\phi(x)\}$ (the factor $a^{-2}$ is multiplied for the
rescaling~(\ref{twoxthirtyfour})). The classical action of the 2d
$\mathcal{N}=(2,2)$ SYM vanishes identically for all configurations of constant
scalar fields such that $[\phi,\overline\phi]=0$ (other fields are set to
zero). These are so-called flat-directions and classical vacua are infinitely
degenerate. Moreover, this degeneracy is not lifted by radiative corrections to
all order of perturbative theory. Thus, the expectation values of scalar fields
in quantum theory are of interest and, if Monte Carlo simulation is useful, a
prediction on these expectation values should be feasible.

First, we consider $a^{-2}\tr\{\phi(x)\overline\phi(x)\}$. This operator is
invariant under the global $\U(1)_R$ transformation, which acts on scalar
fields as $\phi(x)\to e^{2i\alpha}\phi(x)$ and $\overline\phi(x)\to
e^{-2i\alpha}\overline\phi(x)$. The continuum limit of this quantity itself is
meaningless, because it is a bare quantity and suffers from UV divergence. It
should be renormalized. A power counting argument shows that the superficial UV
divergence comes from the simplest one-loop diagram and the divergence is
logarithmic $\sim\ln(a/L)g^2$. If supersymmetry of the 1PI effective action is
restored in the continuum limit, as we assume at the moment, this one-loop
divergence is the only source of UV divergence of
$\llangle a^{-2}\tr\{\phi(x)\overline\phi(x)\}\rrangle$ (recall the argument
below eq.~(\ref{threexseventeen})).

So we define the renormalized operator (the normal product)
\begin{equation}
   \mathcal{N}[a^{-2}\tr\{\phi(x)\overline\phi(x)\}]
   \equiv a^{-2}\tr\{\phi(x)\overline\phi(x)\}-(N_c^2-1)c(a/L)g^2
\label{threextwentythree}
\end{equation}
by subtracting a c-number, the value of the one-loop diagram. This subtraction
must remove all the UV divergence of the composite operator. This simplicity is
a special property of the present two-dimensional (supersymmetric) theory.

The coefficient~$c(a/L)$ of the counter constant is given by a simple scalar
one-loop diagram and, on a finite size lattice, it is
\begin{equation}
   c(a/L=1/N)=\frac{1}{2N^2}\sum_{n_0=0}^{N-1}\sum_{n_1=0}^{N-1}
   \frac{1}{\displaystyle\sum_{\mu=0}^1
   \left(1-\cos\frac{2\pi}{N}n_\mu\right)}.
\label{threextwentyfour}
\end{equation}
As possible prescription for the zero mode, we do not include
$(n_0,n_1)=(0,0)$ in the sum. Values of this counter constant are listed in
table~\ref{table2} for the cases in our simulation.
\TABULAR[t]{|c|c|c|c|c|c|}{\hline
$L/a=N$ & 8 & 7 & 6 & 5 & 4 \\
\hline
$c(a/L)$ & $0.379295$ & $0.357928$ & $0.333234$ & $0.304$ & $0.268229$ \\
\hline}{The counter constant $c(a/L)$ given by
eq.~(\ref{threextwentyfour}).\label{table2}}

The result of our Monte Carlo simulation is figure~\ref{figure9}.\footnote{We
confirmed that the imaginary part is almost negligible (as it should be) and
it is not plotted in the figure.}
\FIGURE[t]{\centerline{\epsfig{file=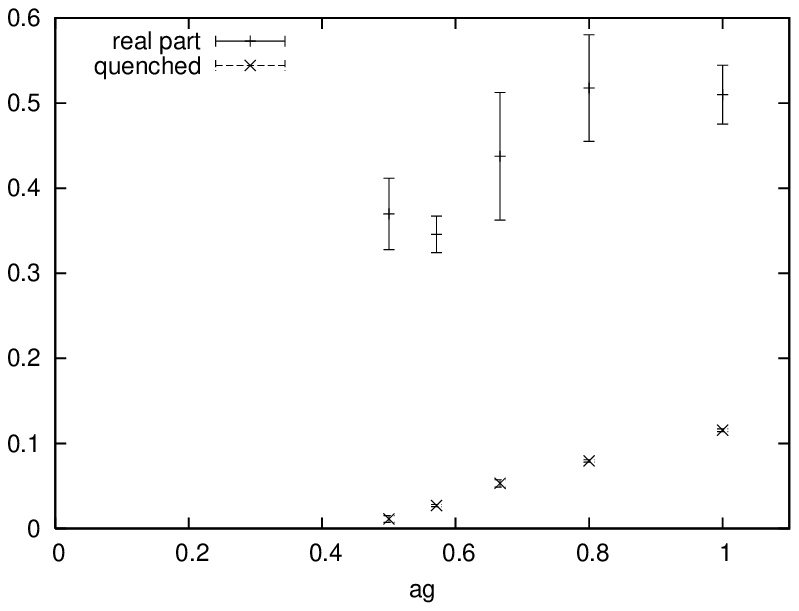,width=0.8\textwidth}}%
\caption{Expectation values of
$\mathcal{N}[a^{-2}\tr\{\phi(x)\overline\phi(x)\}]$.
\label{figure9}}}
First of all, we see clear separation between the re-weighted average and the
quenched one. The difference is thus due to the effect of dynamical fermions.
This effect uplifts the expectation value and this is consistent with the
picture that, in quenched (i.e., non-supersymmetric) theory, the scalar
potential is lifted by radiative corrections, suppressing quantum fluctuation
of scalar fields. As discussed for the WT identity~(\ref{threexseventeen}), if
the supersymmetry is restored in the continuum limit, the expectation value
$\llangle\mathcal{N}[a^{-2}\tr\{\phi(x)\overline\phi(x)\}]\rrangle$ is expected
to become independent of $ag$ as~$a\to0$. The behavior in figure~\ref{figure9}
is more or less consistent with this expectation, although clearly we need
further data at smaller values of~$ag$ to conclude this. In any case,
interestingly, the expectation value appears to approach some finite number (in
a unit of~$g^2$) in the continuum limit after the
renormalization~(\ref{threextwentythree}). (Without the renormalization (the
subtraction), there is a tendency that the expectation values grow as $a\to0$.)
The limiting value of
$\llangle\mathcal{N}[a^{-2}\tr\{\phi(x)\overline\phi(x)\}]\rrangle$ at $a\to0$
(while fixing $Lg$) in the figure itself has no direct physical meaning because
it can freely be shifted by a further finite renormalization. However, the
limiting value should depend on~$Lg$ and this dependence can be a non-trivial
prediction. We need a much finer lattice, of course, for an extrapolation to
the continuum.

In figure~\ref{figure10}, we have plotted
$\llangle a^{-2}\tr\{\phi(x)\phi(x)\}\rrangle$.
\FIGURE[t]{\centerline{\epsfig{file=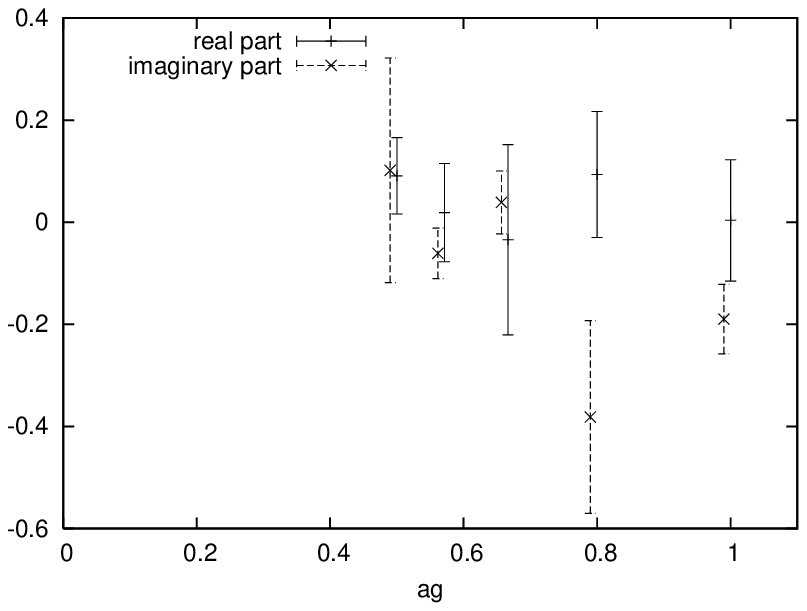,width=0.8\textwidth}}%
\caption{Expectation values of $a^{-2}\tr\{\phi(x)\phi(x)\}$.
\label{figure10}}}
For this, a perturbative argument indicates that there is no need of
renormalization. The result is clearly shows
$\llangle a^{-2}\tr\{\phi(x)\phi(x)\}\rrangle\sim0$. This might be suggested
from the fact that in two dimensions the global $\U(1)_R$ symmetry cannot be
spontaneously broken, although this argument is not rigorous because we are
studying a system in finite volume.

\section{Conclusion}
In this paper, we presented the results of our preliminary numerical study of
Sugino's lattice formulation of the 2d $\mathcal{N}=(2,2)$ SYM. By confirming
WT identities associated with an exact fermionic symmetry of the formulation
in fair accuracy, we infer that the re-weighting method for the dynamical
fermions basically works in this two-dimensional system. On the other hand,
although we could not conclude the restoration of full supersymmetry from
the numerical results, all the results are consistent with the basic idea of
a supersymmetry restoration. We computed also the expectation values of scalar
bi-linear operators to illustrate the usefulness of this kind of lattice
simulation.

In this paper, we did not try to measure any two-point correlation function or
extended observables like Wilson loops, because it is clear that our lattice is
too small to extract any useful information from such quantities. Interesting
physics of this system is, of course, contained in these observables. For
example, the most direct way to examine the restoration (and/or the spontaneous
breaking) of supersymmetry is to study the mass spectra and two-point functions
containing the supersymmetric current. For an interesting property of a
two-point function that contains the $\U(1)_R$ current, see
ref.~\cite{Fukaya:2006mg}. Thanks to FermiQCD/MDP~\cite{DiPierro:2000bd}, our
code is executable also on a large PC cluster without any change. Having
obtained encouraging results in this paper, in the near future, we hope to
report results of full-scale simulation using much larger lattice.

There exists a natural generalization of the present manifestly $Q$-invariant
lattice formulation to the 2d $\mathcal{N}=(4,4)$
SYM~\cite{Sugino:2004qd,Sugino:2004qdf} and to the 2d $\mathcal{N}=(8,8)$
SYM (the second paper of ref.~\cite{Sugino:2003yb}). The latter theory is
especially of interest as an effective theory that describes the dynamics of
D1-brane. We do not find any real difficulty to set up the corresponding Monte
Carlo simulation similar to that of the present paper. This is an interesting
future problem.

\acknowledgments
I would like to thank Hidenori Fukaya, Issaku Kanamori and Tomohisa Takimi
for discussion. I am grateful to Martin L\"uscher for helpful remarks on a
related subject. This work is supported in part by Grant-in-Aid for Scientific
Research (18540305) and by JSPS and French Ministry of Foreign Affairs under
the Japan-France Integrated Action Program (SAKURA).

\ifthenelse{\value{private} = 100}{
%\appendix
%\section{}
\FIGURE[t]{\centerline{\epsfig{file=catterall2i.eps,width=0.8\textwidth}}%
\caption{Expectation values of
$\mathcal{L}_{\text{B}1}(x)+\mathcal{L}_{\text{F}1}(x)$.}}
\FIGURE[t]{\centerline{\epsfig{file=catterall3i.eps,width=0.8\textwidth}}%
\caption{Expectation values of
$2\mathcal{L}_{\text{B}2}(x)+\mathcal{L}_{\text{F}5}(x)$.}}
\FIGURE[t]{\centerline{\epsfig{file=catterall4i.eps,width=0.8\textwidth}}%
\caption{Expectation values of
$\mathcal{L}_{\text{B}3}(x)+\mathcal{L}_{\text{F}3}(x)
+\mathcal{L}_{\text{F}4}(x)+\mathcal{L}_{\text{F}6}(x)$.}}
\FIGURE[t]{\centerline{\epsfig{file=catterall5i.eps,width=0.8\textwidth}}%
\caption{Expectation values of
$\mathcal{L}_{\text{B}1}(x)+\mathcal{L}_{\text{F}2}(x)$.}}
\FIGURE[t]{\centerline{\epsfig{file=catterall6i.eps,width=0.8\textwidth}}%
\caption{Expectation values of
$\mathcal{L}_{\text{B}1}(x)+\mathcal{L}_{\text{F}3}(x)$.}}
}{}

%\appendix

\listoftables           % ONLY DRAFT
\listoffigures          % ONLY DRAFT


\begin{thebibliography}{99}

\bibitem{Sugino:2004qd}
F.~Sugino,
\emph{Super Yang-Mills theories on the two-dimensional lattice with exact
supersymmetry},
\jhep{03}{2004}{067} [\heplat{0401017}].
%%CITATION = HEP-LAT 0401017;%%

\bibitem{Sugino:2004qdf}
F.~Sugino,
\emph{Super Yang-Mills theories on the two-dimensional lattice with exact
supersymmetry}, \heplat{0401017v4}, the latest arXiv version of
ref.~\cite{Sugino:2004qd}.

\bibitem{Curci:1986sm}
G.~Curci and G.~Veneziano,
\emph{Supersymmetry and the lattice: A reconciliation?},
\npb{292}{1987}{555};
%%CITATION = NUPHA,B292,555;%%
\\
%\bibitem{Donini:1997hh}
A.~Donini, M.~Guagnelli, P.~Hernandez and A.~Vladikas,
\emph{Towards $N=1$ super-Yang-Mills on the lattice},
\npb{523}{1998}{529} [\heplat{9710065}];
%%CITATION = NUPHA,B523,529;%%
\\
%\bibitem{Taniguchi:1999fc}
Y.~Taniguchi,
\emph{One loop calculation of SUSY Ward-Takahashi identity on lattice with
Wilson fermion},
\prd{63}{2001}{014502} [\heplat{9906026}];
%%CITATION = HEP-LAT 9906026;%%
\\
%\bibitem{Feo:2003km}
A.~Feo,
\emph{The supersymmetric Ward-Takahashi identity in 1-loop lattice 
perturbation theory. I: General procedure},
\prd{70}{2004}{054504} [\heplat{0305020}].
%%CITATION = HEP-LAT 0305020;%%

\bibitem{Kirchner:1998mp}
R.~Kirchner, I.~Montvay, J.~Westphalen, S.~Luckmann and K.~Spanderen
[DESY-Munster Collaboration],
\emph{Evidence for discrete chiral symmetry breaking in $N=1$ supersymmetric
Yang-Mills theory},
\plb{446}{1999}{209} [\heplat{9810062}].
%%CITATION = PHLTA,B446,209;%%

\bibitem{Campos:1999du}
I.~Campos {\it et al.}  [DESY-Munster Collaboration],
\emph{Monte Carlo simulation of $\SU(2)$ Yang-Mills theory with light
gluinos},
\epjc{11}{1999}{507} [\heplat{9903014}].
%%CITATION = EPHJA,C11,507;%%

\bibitem{Fleming:2000fa}
G.~T.~Fleming, J.~B.~Kogut and P.~M.~Vranas,
\emph{Super Yang-Mills on the lattice with domain wall fermions},
\prd{64}{2001}{034510} [\heplat{0008009}].
%%CITATION = PHRVA,D64,034510;%%

\bibitem{Montvay:2001aj}
I.~Montvay,
\emph{Supersymmetric Yang-Mills theory on the lattice},
\ijmpa{17}{2002}{2377} [\heplat{0112007}],
and references cited therein.
%%CITATION = IMPAE,A17,2377;%%

\bibitem{Farchioni:2001wx}
F.~Farchioni {\it et al.}  [DESY-Munster-Roma Collaboration],
\emph{The supersymmetric Ward identities on the lattice},
\epjc{23}{2002}{719} [\heplat{0111008}].
%%CITATION = EPHJA,C23,719;%%

\bibitem{Farchioni:2004fy}
F.~Farchioni and R.~Peetz,
\emph{The low-lying mass spectrum of the $N=1$ $\SU(2)$ SUSY Yang-Mills theory
with Wilson fermions},
\epjc{39}{2005}{87} [\heplat{0407036}].
%%CITATION = EPHJA,C39,87;%%

\bibitem{Kaplan:2003uh}
D.~B.~Kaplan,
\emph{Recent developments in lattice supersymmetry},
\npps{129}{2004}{109} [\heplat{0309099}].
%%CITATION = NUPHZ,129,109;%%

\bibitem{Feo:2004kx}
A.~Feo,
\emph{Predictions and recent results in SUSY on the lattice},
\mpla{19}{2004}{2387} [\heplat{0410012}].
%%CITATION = MPLAE,A19,2387;%%

\bibitem{Giedt:2006pd}
J.~Giedt,
\emph{Deconstruction and other approaches to supersymmetric lattice field
theories},
\ijmpa{21}{2006}{3039} [\heplat{0602007}];
%%CITATION = IMPAE,A21,3039;%%
%\bibitem{Giedt:2007hz}
%J.~Giedt,
\emph{Advances and applications of lattice supersymmetry},
PoS {\bf LAT2006} (2006) 008 [\heplat{0701006}].
%%CITATION = POSCI,LAT2006,008;%%

\bibitem{Kaplan:2002wv}
D.~B.~Kaplan, E.~Katz and M.~\"Unsal,
\emph{Supersymmetry on a spatial lattice},
\jhep{05}{2003}{037} [\heplat{0206019}];
%%CITATION = HEP-LAT 0206019;%%
\\
%\bibitem{Cohen:2003xe}
A.~G.~Cohen, D.~B.~Kaplan, E.~Katz and M.~\"Unsal,
\emph{Supersymmetry on a Euclidean spacetime lattice. I: A target theory with
four supercharges},
\jhep{08}{2003}{024} [\heplat{0302017}];
%%CITATION = HEP-LAT 0302017;%%
%\bibitem{Cohen:2003qw}
%A.~G.~Cohen, D.~B.~Kaplan, E.~Katz and M.~\"Unsal,
\emph{Supersymmetry on a Euclidean spacetime lattice. II: Target theories
with eight supercharges},
\jhep{12}{2003}{031} [\heplat{0307012}];
%%CITATION = HEP-LAT 0307012;%%
\\
%\bibitem{Kaplan:2005ta}
D.~B.~Kaplan and M.~\"Unsal,
\emph{A Euclidean lattice construction of supersymmetric Yang-Mills theories
with sixteen supercharges},
\jhep{09}{2005}{042} [\heplat{0503039}];
%%CITATION = HEP-LAT 0503039;%%
\\
%\bibitem{Endres:2006ic}
M.~G.~Endres and D.~B.~Kaplan,
\emph{Lattice formulation of $(2,2)$ supersymmetric gauge theories with matter
fields},
\jhep{10}{2006}{076} [\heplat{0604012}].
%%CITATION = JHEPA,0610,076;%%

\bibitem{Giedt:2003ve}
J.~Giedt,
\emph{Non-positive fermion determinants in lattice supersymmetry},
\npb{668}{2003}{138} [\heplat{0304006}];
%%CITATION = NUPHA,B668,138;%%
%\bibitem{Giedt:2003vy}
%J.~Giedt,
\emph{The fermion determinant in $(4,4)$ 2d lattice super-Yang-Mills},
\npb{674}{2003}{259} [\heplat{0307024}];
%%CITATION = NUPHA,B674,259;%%
%\bibitem{Giedt:2003gf}
%J.~Giedt,
\emph{Deconstruction, 2d lattice Yang-Mills, and the dynamical lattice
spacing}, \heplat{0312020};
%%CITATION = HEP-LAT/0312020;%%
%\bibitem{Giedt:2004tn}
%J.~Giedt,
\emph{Deconstruction, 2d lattice super-Yang-Mills, and the dynamical lattice
spacing}, \heplat{0405021}.
%%CITATION = HEP-LAT/0405021;%%

\bibitem{Sugino:2003yb}
F.~Sugino,
\emph{A lattice formulation of super Yang-Mills theories with exact
supersymmetry},
\jhep{01}{2004}{015} [\heplat{0311021}];
%%CITATION = HEP-LAT 0311021;%%
%\bibitem{Sugino:2004uv}
%F.~Sugino,
\emph{Various super Yang-Mills theories with exact supersymmetry on the
lattice},
\jhep{01}{2005}{016} [\heplat{0410035}];
%%CITATION = HEP-LAT 0410035;%%
%\bibitem{Sugino:2006uf}
%F.~Sugino,
\emph{Two-dimensional compact $N=(2,2)$ lattice super Yang-Mills theory with
exact supersymmetry},
\plb{635}{2006}{218} [\heplat{0601024}].
%%CITATION = PHLTA,B635,218;%%

\bibitem{Catterall:2004np}
S.~Catterall,
\emph{A geometrical approach to $N=2$ super Yang-Mills theory on the two 
dimensional lattice},
\jhep{11}{2004}{006} [\heplat{0410052}];
%%CITATION = HEP-LAT 0410052;%%
%\bibitem{Catterall:2005fd}
%S.~Catterall,
\emph{Lattice formulation of $N=4$ super Yang-Mills theory},
\jhep{06}{2005}{027} [\heplat{0503036}].
%%CITATION = HEP-LAT 0503036;%%

\bibitem{Unsal:2005yh}
M.~\"Unsal,
\emph{Compact gauge fields for supersymmetric lattices},
\jhep{11}{2005}{013} [\heplat{0504016}];
%%CITATION = JHEPA,0511,013;%%
%\bibitem{Unsal:2006qp}
%M.~\"Unsal,
\emph{Twisted supersymmetric gauge theories and orbifold lattices},
\jhep{10}{2006}{089} [\hepth{0603046}].
%%CITATION = JHEPA,0610,089;%%

\bibitem{Onogi:2005cz}
T.~Onogi and T.~Takimi,
\emph{Perturbative study of the supersymmetric lattice theory from matrix
model},
\prd{72}{2005}{074504} [\heplat{0506014}].
%%CITATION = PHRVA,D72,074504;%%

\bibitem{Suzuki:2005dx}
H.~Suzuki and Y.~Taniguchi,
\emph{Two-dimensional $\mathcal{N}=(2,2)$ super Yang-Mills theory on
the lattice via dimensional reduction},
\jhep{10}{2005}{082} [\heplat{0507019}].
%%CITATION = HEP-LAT 0507019;%%

\bibitem{D'Adda:2005zk}
A.~D'Adda, I.~Kanamori, N.~Kawamoto and K.~Nagata,
\emph{Exact extended supersymmetry on a lattice: Twisted $N=2$ super Yang-Mills
in two dimensions},
\plb{633}{2006}{645} [\heplat{0507029}].
%%CITATION = PHLTA,B633,645;%%

\bibitem{Elliott:2005bd}
J.~W.~Elliott and G.~D.~Moore,
\emph{Three dimensional $N=2$ supersymmetry on the lattice},
PoS {\bf LAT2005} (2006) 245 [\jhep{11}{2005}{010}]
[\heplat{0509032}].
%%CITATION = JHEPA,0511,010;%%

\bibitem{Catterall:2006jw}
S.~Catterall,
\emph{Simulations of $\mathcal{N}=2$ super Yang-Mills theory in two
dimensions},
\jhep{03}{2006}{032} [\heplat{0602004}];
%%CITATION = HEP-LAT 0602004;%%
%\bibitem{Catterall:2006is}
%S.~Catterall,
\emph{On the restoration of supersymmetry in twisted two-dimensional lattice
Yang-Mills theory},
\jhep{04}{2007}{015} [\heplat{0612008}].
%%CITATION = HEP-LAT 0612008;%%

\bibitem{Ohta:2006qz}
K.~Ohta and T.~Takimi,
\emph{Lattice formulation of two dimensional topological field theory},
\ptp{117}{2007}{317} [\heplat{0611011}].
%%CITATION = PTPKA,117,317;%%

\bibitem{Damgaard:2007be}
P.~H.~Damgaard and S.~Matsuura,
\emph{Classification of supersymmetric lattice gauge theories by orbifolding},
\jhep{07}{2007}{051} [{\tt arXiv:0704.2696} [{\tt hep-lat}]].
%%CITATION = JHEPA,0707,051;%%

\bibitem{Takimi:2007nn}
T.~Takimi,
\emph{Relationship between various supersymmetric lattice models},
\jhep{07}{2007}{010} [{\tt arXiv:0705.3831} [{\tt hep-lat}]].
%%CITATION = JHEPA,0707,010;%%

\bibitem{Matsumura:1995kw}
Y.~Matsumura, N.~Sakai and T.~Sakai,
\emph{Mass spectra of supersymmetric Yang-Mills theories in $(1+1)$-dimensions},
\prd{52}{1995}{2446} [\hepth{9504150}];
%%CITATION = PHRVA,D52,2446;%%
\\
%\bibitem{Antonuccio:1998tm}
F.~Antonuccio, O.~Lunin, S.~Pinsky, H.~C.~Pauli and S.~Tsujimaru,
\emph{The DLCQ spectrum of $N=(8,8)$ super Yang-Mills},
\prd{58}{1998}{105024} [\hepth{9806133}];
%%CITATION = PHRVA,D58,105024;%%
\\
%\bibitem{Antonuccio:1998mq}
F.~Antonuccio, H.~C.~Pauli, S.~Pinsky and S.~Tsujimaru,
\emph{DLCQ bound states of $N=(2,2)$ super-Yang-Mills at finite and large N},
\prd{58}{1998}{125006} [\hepth{9808120}];
%%CITATION = PHRVA,D58,125006;%%
\\
%\bibitem{Antonuccio:1999iz}
F.~Antonuccio, A.~Hashimoto, O.~Lunin and S.~Pinsky,
\emph{Can DLCQ test the Maldacena conjecture?},
\jhep{07}{1999}{029} [\hepth{9906087}];
%%CITATION = JHEPA,9907,029;%%
\\
%\bibitem{Hiller:2000nf}
J.~R.~Hiller, O.~Lunin, S.~Pinsky and U.~Trittmann,
\emph{Towards a SDLCQ test of the Maldacena conjecture},
\plb{482}{2000}{409} [\hepth{0003249}];
%%CITATION = PHLTA,B482,409;%%
\\
%\bibitem{Hiller:2003qe}
J.~R.~Hiller, S.~S.~Pinsky and U.~Trittmann,
\emph{Anomalously light mesons in a $(1+1)$-dimensional supersymmetric theory
with fundamental matter},
\npb{661}{2003}{99} [\hepph{0302119}];
%%CITATION = NUPHA,B661,99;%%
\\
%\bibitem{Harada:2004ck}
M.~Harada, J.~R.~Hiller, S.~Pinsky and N.~Salwen,
\emph{Improved results for $N=(2,2)$ super Yang-Mills theory using
supersymmetric discrete light-cone quantization},
\prd{70}{2004}{045015} [\hepth{0404123}];
%%CITATION = PHRVA,D70,045015;%%
\\
%\bibitem{Hiller:2005vf}
J.~R.~Hiller, S.~S.~Pinsky, N.~Salwen and U.~Trittmann,
\emph{Direct evidence for the Maldacena conjecture for $N=(8,8)$ super
Yang-Mills theory in $1+1$ dimensions},
\plb{624}{2005}{105} [\hepth{0506225}].
%%CITATION = PHLTA,B624,105;%%

\bibitem{Witten:1988ze}
E.~Witten,
\emph{Topological quantum field theory},
\cmp{117}{1988}{353};
%%CITATION = CMPHA,117,353;%%
%\bibitem{Witten:1990bs}
%E.~Witten,
\emph{Introduction to cohomological field theories},
\ijmpa{6}{1991}{2775}.
%%CITATION = IMPAE,A6,2775;%%

\bibitem{Catterall:2001fr}
S.~Catterall and S.~Karamov,
\emph{Exact lattice supersymmetry: the two-dimensional $N=2$ Wess-Zumino
model},
\prd{65}{2002}{094501} [\heplat{0108024}];
%CITATION = HEP-LAT 0108024;%%
\\
%\bibitem{Catterall:2003wd}
S.~Catterall,
\emph{Lattice supersymmetry and topological field theory},
\jhep{05}{2003}{038} [\heplat{0301028}].
%%CITATION = HEP-LAT 0301028;%%

\bibitem{Elitzur:1982vh}
S.~Elitzur, E.~Rabinovici and A.~Schwimmer,
\emph{Supersymmetric models on the lattice},
\plb{119}{1982}{165}.
%%CITATION = PHLTA,B119,165;%%

\bibitem{Luscher:1999du}
M.~L\"uscher,
\emph{Abelian chiral gauge theories on the lattice with exact gauge
invariance},
\npb{549}{1999}{295} [\heplat{9811032}].
%%CITATION = HEP-LAT 9811032;%%

\bibitem{Durr:2003xs}
S.~D\"urr and C.~Hoelbling,
\emph{Staggered versus overlap fermions: A study in the Schwinger model with
$N_f=0$, 1, 2},
\prd{69}{2004}{034503} [\heplat{0311002}].
%%CITATION = PHRVA,D69,034503;%%

\bibitem{Montvay:1994cy}
I.~Montvay and G.~M\"unster,
\emph{Quantum fields on a lattice},
Cambridge University Press (1994).
%\href{http://www.slac.stanford.edu/spires/find/hep/www?irn=3030474}{SPIRES entry}

\bibitem{DiPierro:2000bd}
M.~Di Pierro,
\emph{Matrix Distributed Processing:
A set of C++ Tools for implementing generic lattice computations on parallel
systems},
\cpc{141}{2001}{98} [\heplat{0004007}];
\\
%%CITATION = CPHCB,141,98;%%
%\bibitem{DiPierro:2005qx}
M.~Di Pierro and J.~M.~Flynn,
\emph{Lattice QFT with FermiQCD},
{\it PoS\/} {\bf LAT2005} (2006) 104 [\heplat{0509058}].
%%CITATION = POSCI,LAT2005,104;%%

\bibitem{Fukaya:2006mg}
H.~Fukaya, M.~Hayakawa, I.~Kanamori, H.~Suzuki and T.~Takimi,
\emph{Note on massless bosonic states in two-dimensional field theories},
\ptp{116}{2007}{1117} [\hepth{0609049}].
%%CITATION = PTPKA,116,1117;%%

\end{thebibliography}
\end{document}